\documentclass[10pt]{article}

\usepackage{graphicx}
\usepackage[colorinlistoftodos]{todonotes}
\usepackage{amsfonts}
\usepackage{amsmath}
\usepackage{subcaption}
\usepackage{dsfont}
\usepackage{mathrsfs}
\usepackage{amsthm}
\newtheorem*{theorem}{Theorem}

\newcommand{\opr}[1]{\ensuremath{\widehat{\mathrm{#1}}}}

\newcommand{\decspace}{\!\!\!\!\!\!\!\!\!\!}
\begin{document}
	
	\title{Dirac delta-convergence of free-motion time-of-arrival eigenfunctions}
	\author{John Jaykel P. Magadan\thanks{jpmagadan@up.edu.ph } $\>$ and Eric A. Galapon\\Theoretical Physics Group, National Institute of Physics\\University of the Philippines, 1101 Philippines}

	\maketitle
	\begin{abstract}
		Previous numerical analyses on the Aharonov-Bohm (AB) operator representing the quantum time-of-arrival (TOA) observable for the free particle have indicated that its eigenfunctions represent quantum states with definite arrival time at the arrival point. In this paper, we give the mathematical proof that this is indeed the case. An essential element of this proof is the consideration of the eigenfunctions of the AB operator with complex eigenvalues.  These eigenfunctions can be considered legitimate TOA eigenfunctions because they evolve unitarily to collapse at the arrival point at the time equal to the real part of their eigenvalue. We show that the time-evolved TOA position probability density distribution evaluated at the time equal to the real part of the eigenvalue forms a dirac delta sequence in the limit as the imaginary part of the eigenvalue approaches zero.
	\end{abstract}
	
	\section{Introduction}
	
	Despite the immense success of quantum theory, its inability to explain temporal-related events is an important gap that needs to be addressed. A feature of time that is of current interest is the quantum time-of-arrival (TOA) problem, which is the measurement of time at which a quantum particle will arrive at a specific location. Classically, the particle's TOA is well-defined because of its definite trajectories. Quantum mechanically, the probabilistic nature of the quantum measurement makes the TOA description statistical in nature. Several prescriptions have been proposed on how to determine the arrival time distribution of a quantum particle, e.g. semi-classical approach \cite{muga2}, operational methods \cite{muga2}, Bohm's trajectory approach \cite{muga2,leavens}, operator method \cite{Busch_pla,toa4,giannitrapani1997,delgado1997PRA,delgado1998PRA,leon1997JPA,Muga1998,mug21,kijowski1999PRA,caballar2,galaponJMP2004,galaponPRA2005a,galaponPRA2005b,villanuevaJPA2008,villanuevaPRA2010,galaponPRA2009,galaponPRSA2009,sombilloJMP2012,sombilloAP2016,EAGJJM2018,DAPJJMCAEAG2024}, and each of these prescriptions gives a unique prediction on the particle's arrival time distribution.  Experiments such as the electron time-of-flight in a cylindrical waveguide \cite{das_SciRep_2019,stopp_NJPhys_2021} and the double-slit experiment \cite{ayatollah_CommunPhys_2023,ayatollah_SciRep_2024,das_AnnPhys_2025} have been proposed to determine which of these prescriptions is consistent with the experimental results. Resolving the QTOA problem is important because time-of-flight experiments have long been performed and TOF data are being used in computations crucial to particle physics experiments \cite{galaponPRA2009}. In addition, the quantum tunneling time problem can be analyzed as a quantum TOA problem, and recent works have shown that the TOA operator approach can explain the zero tunneling time results that are observed in attosecond tunneling time experiments \cite{galaponPRL2012,DAPEAG2020,PCFDAPEAG2024_2,PCFDAPEAG2024_1,Sainadh2019}.
	
	The standard approach to representing observables in quantum mechanics is to represent them by operators, and these operators are usually constructed by quantizing their classical counterparts. For a free particle with mass $\mu$, the classical arrival time at the origin ($X=0$) is $T(q,p) = -\mu q/p$ and its symmetric quantization is given by
	\begin{equation}
	\opr{T} = -\frac{\mu}{2}(\opr{p}^{-1}\opr{q} + \opr{q}\opr{p}^{-1}).
	\end{equation}
	This operator is referred to as the Aharonov-Bohm (AB) operator, since it is the negative of the operator introduced by Aharonov and Bohm as a quantum clock to measure the duration in which the energy of the particle is measured \cite{Aharonov1961}.
	Much is known about the mathematical properties of $\opr{T}$ when defined in the entire real line \cite{paul1962}. It is maximally symmetric and canonically conjugate with the free particle hamiltonian $\opr{H} = \opr{p}^2/2\mu$. It has two degenerate generalized eigenfunctions which satisfy covariance under time translation and form a complete set. Its associated arrival-time distribution is the Kijowski's distribution, which is considered the ideal arrival time distribution for a free particle \cite{Muga1998, kijowski}.
	
	An important requirement for a legitimate quantum TOA operator is that its eigenfunctions should describe a state with definite arrival time. This means that the particle, when prepared as a TOA eigenstate, should be located exactly at the arrival point at a time equal to its corresponding eigenvalue. Muga et al. investigated whether the AB operator exhibits such a property by examining the time-evolved form of the eigenfunctions in position representation \cite{Muga1998}. The issue of non-normalizability of the eigenfunctions was addressed by constructing a normalized wave packet using a scaling factor on the TOA eigenfunction. It was observed that the normalized wave packets evolve according to the Schrodinger equation such that the average position equals the arrival point at the time equal to the corresponding eigenvalue $T$, which is also the time when the spatial width of the wave packet is at minimum. In the limit as the scaling factor approaches unity, the wave packet's position probability density becomes more and more peaked at the space-time point $(x=0,t=T)$, indicating that these normalized wavepackets describe a quantum particle that will arrive at the arrival point at a time equal to its respective eigenvalue. 
	
	Another approach used to investigate the dynamical property of $\opr{T}$ is by spatial confinement. The resulting operator, called the confined time-of-arrival (CTOA) operator, is a self-adjoint operator with a discrete eigenvalue spectrum and doubly degenerate normalizable eigenfunctions \cite{galaponPRA2005a}. The eigenfunctions form a complete set of basis. However, they do not exhibit covariance under time translation because of their discrete spectrum. The normalizability of the CTOA eigenfunctions allows the investigation of their dynamics in the position representation without the need of scaling factors. It was shown that CTOA eigenfunctions evolve according to the Schrodinger equation `such that the event at which its position expectation value equals the arrival point and the event at which its position variance is at minimum simultaneously occur at a later time equal their corresponding eigenvalues' \cite{caballar2, galaponPRA2005a}. This dynamical behaviour is consistent with that of the normalized TOA wave packets in \cite{Muga1998}, and this dynamical property is referred to as the unitary arrival or unitary collapse to distinguish it from the wavefunction collapse that results from observable measurement. The collapse becomes sharper as $l$ increases \cite{galaponPRA2005a, galaponPRA2005b}, indicating that the TOA probability density approaches a function with singular support as $l\to \infty$. In addition, the CTOA distribution approaches the Kijowski distribution as  $l\to\infty$ which shows the connection of the CTOA operator with AB operator defined in the entire real line \cite{galaponPRA2005b}.
	
	Although the results in \cite{Muga1998} and \cite{galaponPRA2005b} seem to indicate that the limiting case, i.e. generalized eigenfunctions of the AB operator, describe states with definite arrival time at the arrival point, they still do not give a valid proof that it is indeed the case because their analysis and conclusions were based on numerical results. Furthermore, the introduction of the scaling factor in \cite{Muga1998} is an ad hoc procedure, and normalized wavepackets do not satisfy the TOA eigenvalue equation. Hence, it is the objective of this paper to give a formal proof that the generalized eigenfunctions of the AB operator are quantum states with definite arrival times at the arrival point. We will show using a theorem for a dirac delta sequence that these eigenfunctions evolve unitarily so that the position probability density at the time equal to their corresponding eigenvalues is a dirac delta function with the arrival point as its support. We will also see that different TOA eigenfunction types have corresponding distinct delta sequences. One of these is an example of the delta sequence first shown in \cite{galaponJPA2009} where all elements in the sequence vanish at the support of the limit dirac delta function.
	
	A crucial element of this proof is the consideration of the eigenfunctions of the AB operator with complex eigenvalues. Previous works on $\opr{T}$ do not consider these states as its eigenfunctions because they reside outside the domain of the hermitian version of $\opr{T}$. However, these are normalizable wavefunctions and may represent physically realizable states. The consideration of eigenfunctions with complex eigenvalues is not a new concept in quantum mechanics. The coherent states, which are the complex-eigenvalued eigenfunctions of the annihilation operator, have an important role in quantum mechanics, quantum optics and quantum electrodynamics. The real and imaginary parts of the eigenvalue are physically meaningful; the real part tells us the state's position expectation value, while the imaginary part tells us the state's momentum expectation value. In the same manner, we will see that both the real and imaginary parts of the complex eigenvalues of the AB operator are also physically meaningful in the context of quantum TOA observable. 
	
	The remainder of the paper is organized as follows. In Section 2, we discuss the eigenfunctions of the AB operator when defined in the entire real line. In Section 3, we discuss the dynamical behavior of these eigenfunctions and the physical significance of the real and imaginary parts of the eigenvalues. In Section 4, we give the proof using a theorem on delta sequence representation that the generalized eigenfunctions of the AB operator are states with definite arrival times at the arrival point.
	
\section{Complex-eigenvalued eigenfunctions of the AB operator}\label{S-TOA}
A quantum particle moving freely in space has an associated Hilbert space $\mathcal{H} = L^2(\mathcal{R},\mathrm{d}p).$ In momentum representation, the AB operator assumes the form
\begin{equation}\label{eq:toa_eigeg_p}
\opr{T}\phi(p) = -i\mu\hbar \left(\frac{1}{p}\frac{\partial}{\partial p} - \frac{1}{2p^2}\right)\phi(p)
\end{equation}
To fully define $\opr{T}$, its domain in $\mathcal{H}$ must be specified and this depends on the properties that $\opr{T}$ must satisfy. In principle, standard quantum mechanics requires the operator representing an observable to be self-adjoint. However, for the case of $\opr{T}$ defined in $\mathcal{H}$, it is maximally hermitian, implying that there is no subspace in $\mathcal{H}$ such that $\opr{T}$ is self-adjoint \cite{mug5}. The maximal domain $D_{max}(\opr{T})$ of $\opr{T}$, which is the largest domain that $\opr{T}$ can admit, consists of all $\phi \in \mathcal{H}$ such that $\opr{T}\phi \in \mathcal{H}$. In this domain, $\opr{T}$ is non-hermitian. In contrast, the domain in which $\opr{T}$ is hermitian consists of all $\psi(p)\in D_{max}(\opr{T})$ that also satisfy the condition $\psi(p)/p^{3/2}\to 0 $ as $|p|\to 0$. We denote the hermitian version of $\opr{T}$ as $\opr{T}_h$. Based on the conditions for these domains, we see that $D(\opr{T}_h) \subset D_{max}(\opr{T})\subset \mathcal{H}$ and that $\opr{T}$ can be hermitian or non-hermitian depending on its associated domain.

The degenerate eigenfunction solution for $\opr{T}$ is given by	
\begin{equation}\label{eq:toa_peigf3}
\phi^{\alpha}_{\tau} (p) = \sqrt{|p|}\exp\left(\frac{i p^2 \tau}{2\mu\hbar}\right) \Theta(\alpha p)
\end{equation}
where $\tau$ is the corresponding eigenvalue, $\alpha = \pm$ is the sign of momentum, and $ \Theta(x)$ is the Heaviside function. Mathematically, $\tau$ can be complex-valued. The eigenfunction $\phi^{\alpha}_{\tau}(p)$ is normalizable for $Im(\tau)>0$ and non-normalizable for $Im(\tau)\leq 0$ so that $\phi^{\alpha}_{\tau}(p) \in D_{max}(\opr{T})$ for $Im(\tau)>0$ and are eigenfunctions of non-hermitan $\opr{T}$. However, they do not satisfy the condition for the elements of $D(\opr{T}_h)$ so they are not eigenfunctions of $\opr{T}_h$. Although  $\opr{T}_h$ has no eigenfunctions in $\mathcal{H}$, it accommodates $\phi^{\alpha}_{\tau}(p)$ for real $\tau$ as its generalized eigenfunctions. Earlier works on the free TOA observable require $\opr{T}$ to be hermitian. As a result, the eigenfunctions $\phi^{\alpha}_{\tau}(p)$ for $Im(\tau)>0$ are often neglected in the investigation of quantum free TOA \cite{paul1962}. However, these are normalizable wavefunctions that may represent physically realizable quantum states, so it is meaningful to investigate their properties and their significance in relation to quantum TOA observables. 

Now, we obtain the eigenfunction's position representation. First, we write the eigenfunctions in the odd-even form $\varphi_{\tau, n=0,1} (p) = (\phi^{+}_{\tau}(p) + (-1)^n \phi^{-}_{\tau}(p))/\sqrt{2}$ where $n =0,1$ with $n=0$ corresponding to the even form, while $n=1$ corresponding to the odd form. These eigenfunctions corresponds to the nodal-nonnodal eigenfunction type of the CTOA operator \cite{sombilloAP2016}. In their normalized form, they are given by
\begin{equation}\label{eq:toa_peigf1}
\varphi_{\tau, 0} (p) = \sqrt{\frac{|p|\tau_I}{\mu\hbar}} \exp\left(\frac{i p^2 \tau}{2\mu\hbar}\right)
\end{equation}    
\begin{equation}\label{eq:toa_peigf2}
\varphi_{\tau,1} (p) = \sqrt{\frac{|p|\tau_I}{\mu\hbar}} \exp\left(\frac{i p^2 \tau} {2\mu\hbar}\right)\mbox{sgn}(p)
\end{equation}    
To obtain  $\varphi_{\tau, n}(q)$, we use the fourier transform formula 
\begin{equation}
\int_{-\infty}^{\infty} \mathrm{d}x \> e^{-ikx} \>_1F_1\left(a;b;- x^2\right) = \frac{\sqrt{\pi}\Gamma(b)}{\Gamma(a)}k^{2a-1}e^{-k^2}U\left(b-\frac{1}{2};a+\frac{1}{2};k^2\right)
\end{equation}
where $U(\alpha;\beta;z)$ is the confluent hypergeometric function of the second kind. For specific values of $a$ and $b$ and by shifting the contour integration in the complex plane, we obtain the following formulas
\begin{equation}\label{eq:ft1}
\int_{-\infty}^{\infty} \mathrm{d}x \> e^{-ikx} \>_1F_1\left(\frac{3}{4};\frac{1}{2};-i\beta x^2\right) = \frac{2^{3/2}(-1)^{5/8}\pi \sqrt{|k|}}{\beta^{3/4}\Gamma(-\frac{1}{4})}e^{ik^2/4\beta}
\end{equation}
\begin{equation}\label{eq:ft2}
\int_{-\infty}^{\infty} \mathrm{d}x \> e^{-ikx} x \>_1F_1\left(\frac{5}{4};\frac{3}{2};-i\beta x^2\right) = \frac{(-1)^{7/8}\pi \sqrt{2|k|}}{\beta^{5/4}\Gamma(1/4)}e^{ik^2/4\beta}\mbox{sgn}(k),
\end{equation}
which hold for $Im(\beta) <0$. Using inverse fourier transform on Eqs. \eqref{eq:ft1} and \eqref{eq:ft2}, we obtain
\begin{equation}\label{eq:AB_eigf_pos}
\varphi_{\tau,n}(q) = \sqrt{\frac{8\tau_I \pi}{|\tau|}}\>\Gamma\left(\frac{1}{4}+\frac{n}{2}\right)^{-1}\left[\frac{\mu}{8\hbar |\tau|}\right]^{\frac{1}{4}+\frac{n}{2}}q^n \>_1F_1\left(\frac{3}{4}+\frac{n}{2};\frac{1}{2}+n;-\frac{i\mu q^2}{2\hbar \tau}\right)
\end{equation}
We now investigate the asymptotic behaviour of these eigenfunctions for large $q$ to confirm the conditions on $\tau$ at which $\varphi_{\tau,n}(q)$ is normalizable. Using the asymptotic expansion of the hypergeometric function $\>_1F_1 (a;b;z)$ as $|z|\to\infty$ \cite{wolfram1F1}
\begin{eqnarray}
\>_1F_1(a;b;z) &\propto& \Gamma(b)\left[\frac{e^z z^{a-b}}{\Gamma(a)}\left(1 + \frac{(a-1)(a-b)}{z}+ \cdots\right)\right. \nonumber \\
&+&\left. \frac{(-z)^{-a}}{\Gamma(b-a)}\left( 1-\frac{a(a-b+1)}{z}+\cdots\right)\right], \qquad |z| \to \infty,
\end{eqnarray}
we see that $\varphi_{\tau,n}(q)$ behaves as
\begin{eqnarray}\label{eq:toa_eigf_asymp}
\varphi_{\tau,n}(q) &\propto& C_n\frac{\Gamma\left(\frac{1}{2}+n\right)}{\Gamma\left(-\frac{1}{4}+\frac{n}{2}\right)}\left[\frac{4 q^{1/2}}{2n-1}\left(\frac{-i\mu }{2\hbar \tau}\right)^{1/4-n/2}e^{-\frac{i\mu q^2}{2\hbar \tau}}\left(1 + O(q^{-2})\right) \right. \nonumber \\ 
&+& \left. q^{-3/2}\left(\frac{i\mu }{2\hbar \tau}\right)^{-3/4-n/2}\left(1 + O(q^{-2})\right)\right], \qquad |q|\to \infty
\end{eqnarray}
where $C_n =  \sqrt{\frac{8\tau_I \pi}{|\tau|}}\>\Gamma\left(\frac{1}{4}+\frac{n}{2}\right)^{-1}\left[\frac{\mu}{8\hbar |\tau|}\right]^{\frac{1}{4}+\frac{n}{2}}$. Let $\tau = \tau_R + i\tau_I$ where $\tau_R$ and $\tau_I$ are the real and imaginary parts of $\tau$. As $|q|\to\infty$, the second term in Eq. \eqref{eq:toa_eigf_asymp} behaves similar  to $O(q^{-3/2})$ while that of the first term depends on the value of $\tau_I$. For $\tau_I>0$, the first term in \eqref{eq:toa_eigf_asymp} decays exponentially as $|q|\to \infty$, so that $\varphi_{\tau,n}(q)\sim O(|q|^{-3/2})$ as $|q|\to\infty$. For $\tau_I<0$, the first term grows exponentially as $|q|\to\infty$, so that $\varphi_{\tau,n}(q)$ is nonnormalizable. Lastly, for real $\tau$, $\varphi_{\tau,n}(q)\sim |q|^{1/2}e^{-\frac{i\mu q^2}{2\hbar \tau_R}}$ as $|q|\to \infty$ so that $\varphi_{\tau,n}(q)$ is also non-normalizable for real $\tau$. Hence, $\varphi_{\tau,n}(q)$ is normalizable only for $\tau_I > 0$, consistent to the result we obtained above. 
	
	\section{Physical significance of complex-eigenvalued eigenfunctions of the AB operator}
	
	We now investigate the dynamical properties of $\varphi_{\tau,n}(q)$ for $\tau_I>0$. Our goal here is to determine if these eigenfunctions can be considered as TOA eigenfunctions by examining whether they exhibit the unitary arrival property. We also aim to determine the physical significance of the real and imaginary parts of $\tau$.
	
\begin{figure}[t!]	
		\centering
		\begin{subfigure}[b]{0.47\textwidth}
			\includegraphics[width=\textwidth]{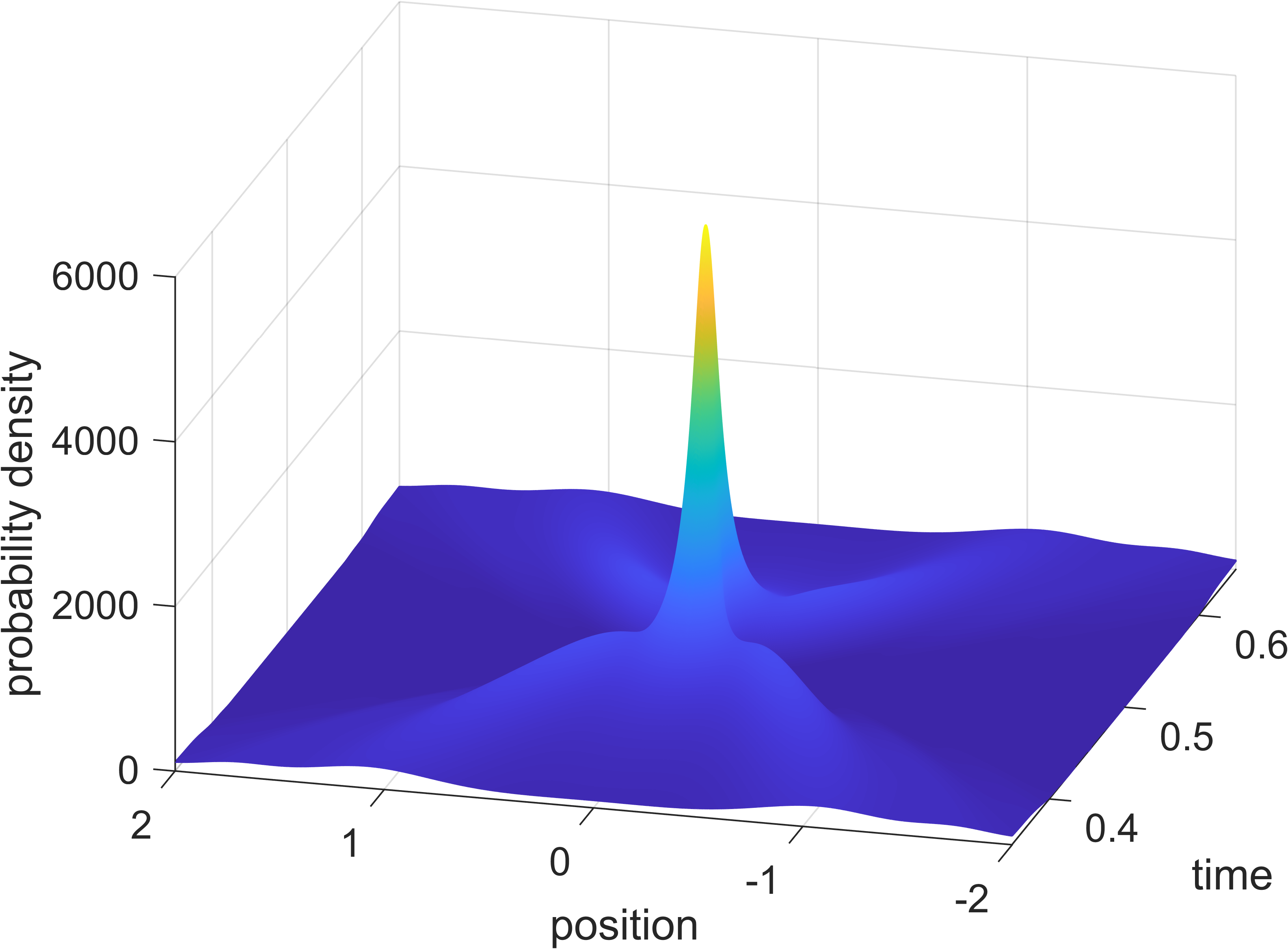}	   
			\caption{}
			\label{fig:nonnodal_b1}
		\end{subfigure}
		\begin{subfigure}[b]{0.47\textwidth}
			\includegraphics[width=\textwidth]{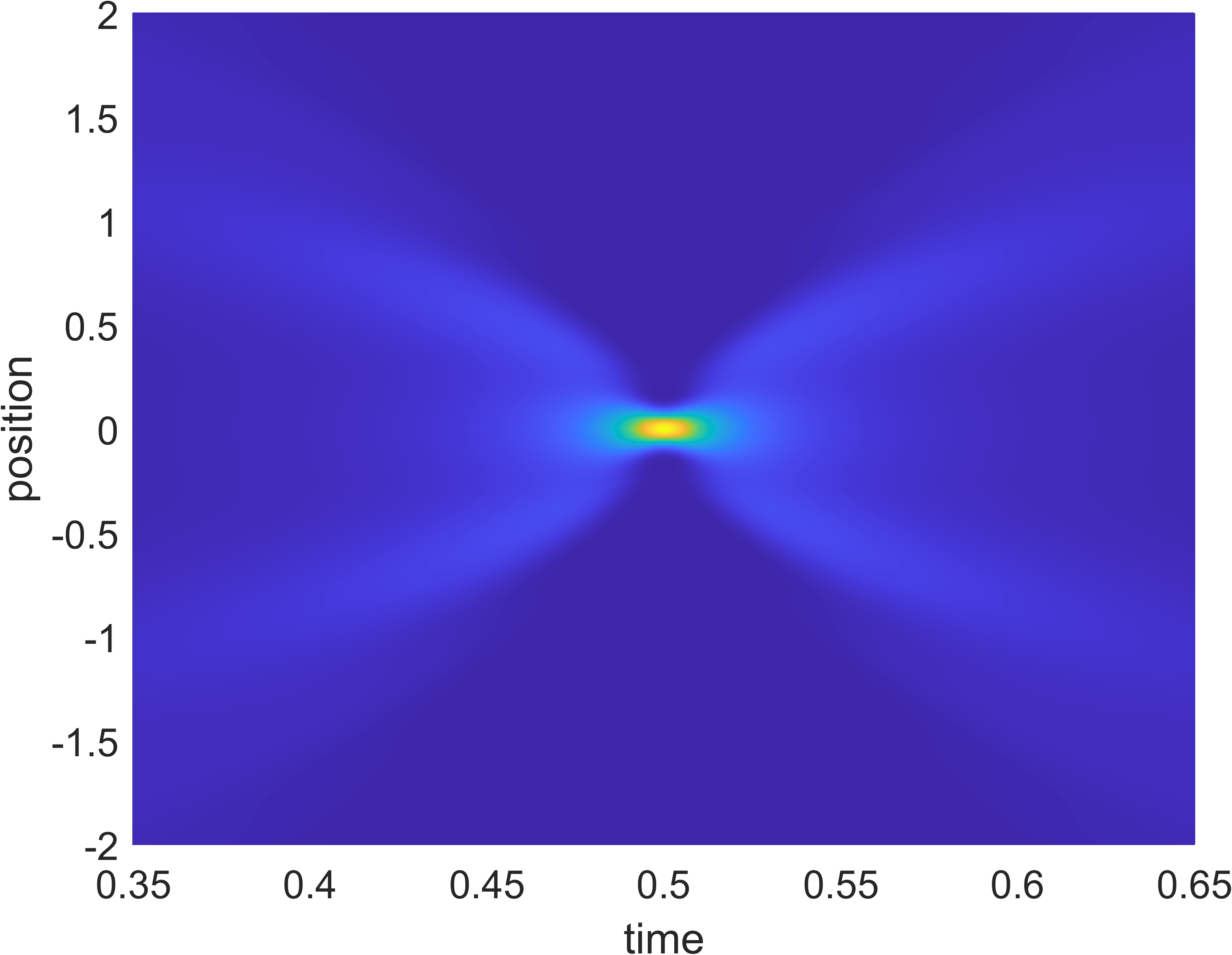}
			\caption{}
			\label{fig:nonnodal_top_b1}
		\end{subfigure}
		\begin{subfigure}[b]{0.47\textwidth}
			\includegraphics[width=\textwidth]{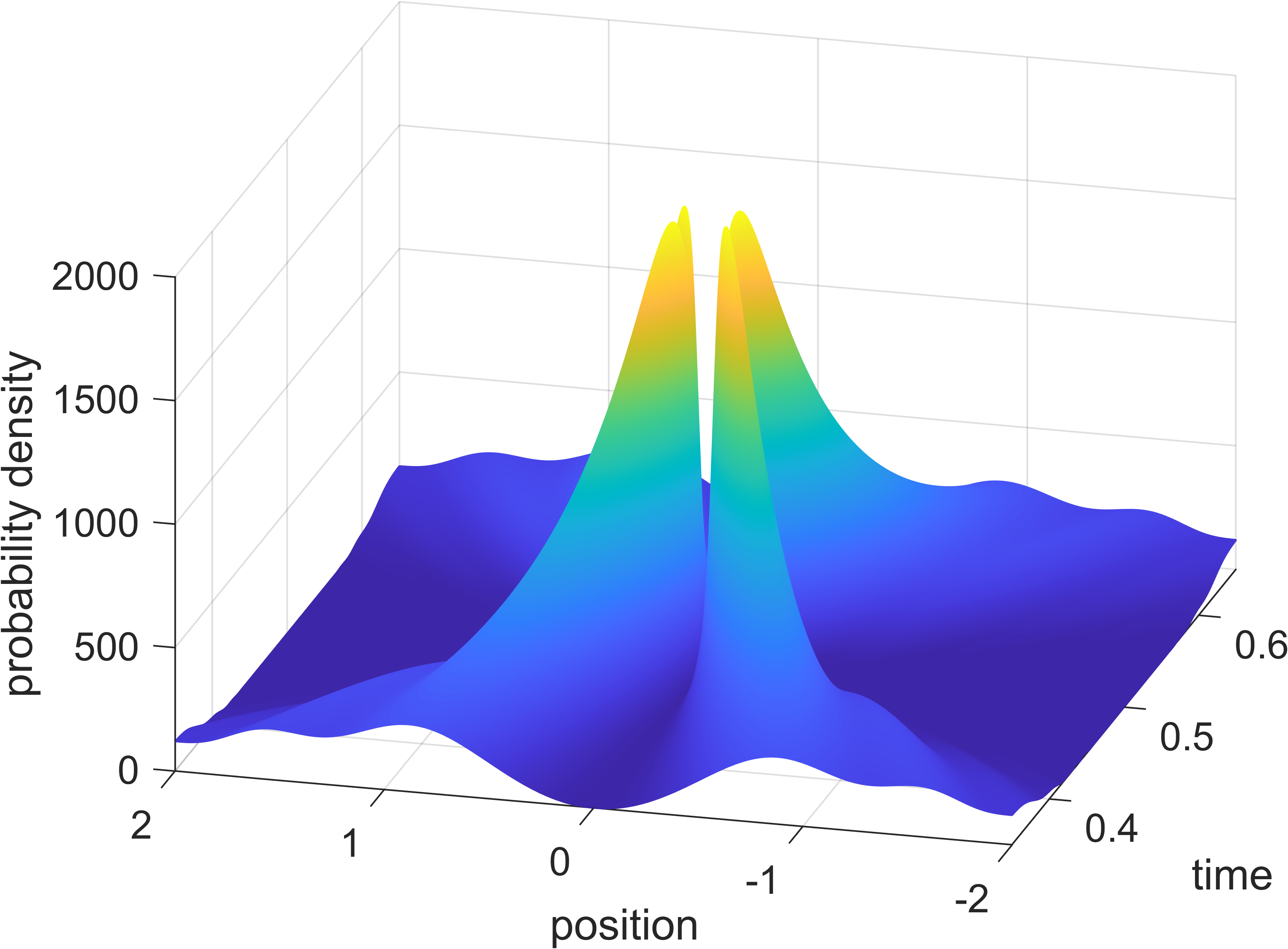}
			\caption{}
			\label{fig:nodal_b1}
		\end{subfigure}
		\begin{subfigure}[b]{0.47\textwidth}
			\includegraphics[width=\textwidth]{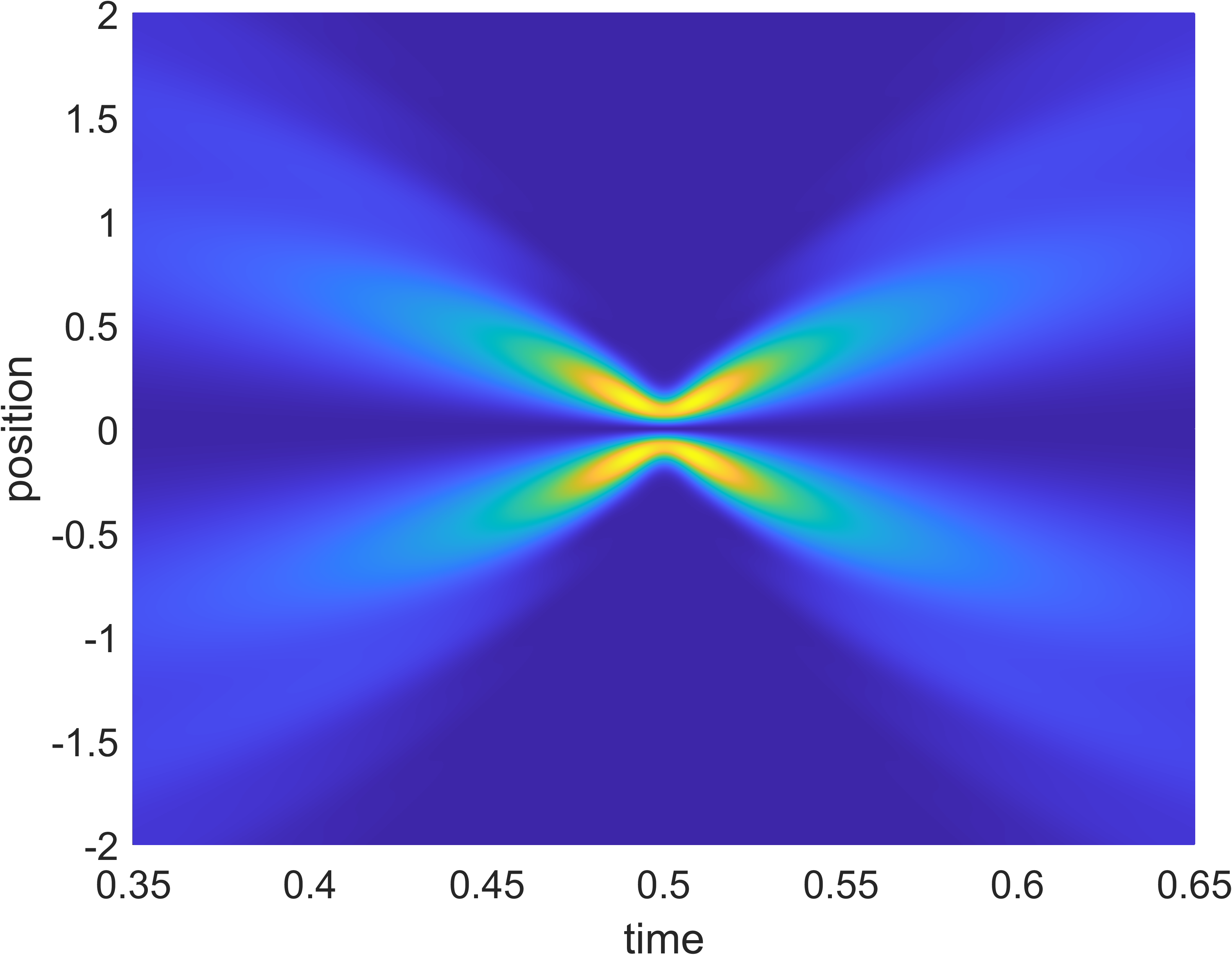}
			\caption{}
			\label{fig:nodal_top_b1}
		\end{subfigure}
		\caption{Time-evolved probability densities for the eigenfunctions  $\varphi_{\tau,0}(q,t)$ $(a,b)$ and  $\varphi_{\tau,1}(q,t)$ $(c,d)$ of the AB operator with $\tau = 0.5 + 0.01i$. The eigenfunctions exhibit unitary collapse with the time at which the collapse is sharpest occurring at $t=\tau_R$. }
		\label{fig:TOAevolve-cmplx_t1}
	\end{figure}

\begin{figure}[t!]	
		\centering
		\begin{subfigure}[b]{0.47\textwidth}
			\includegraphics[width=\textwidth]{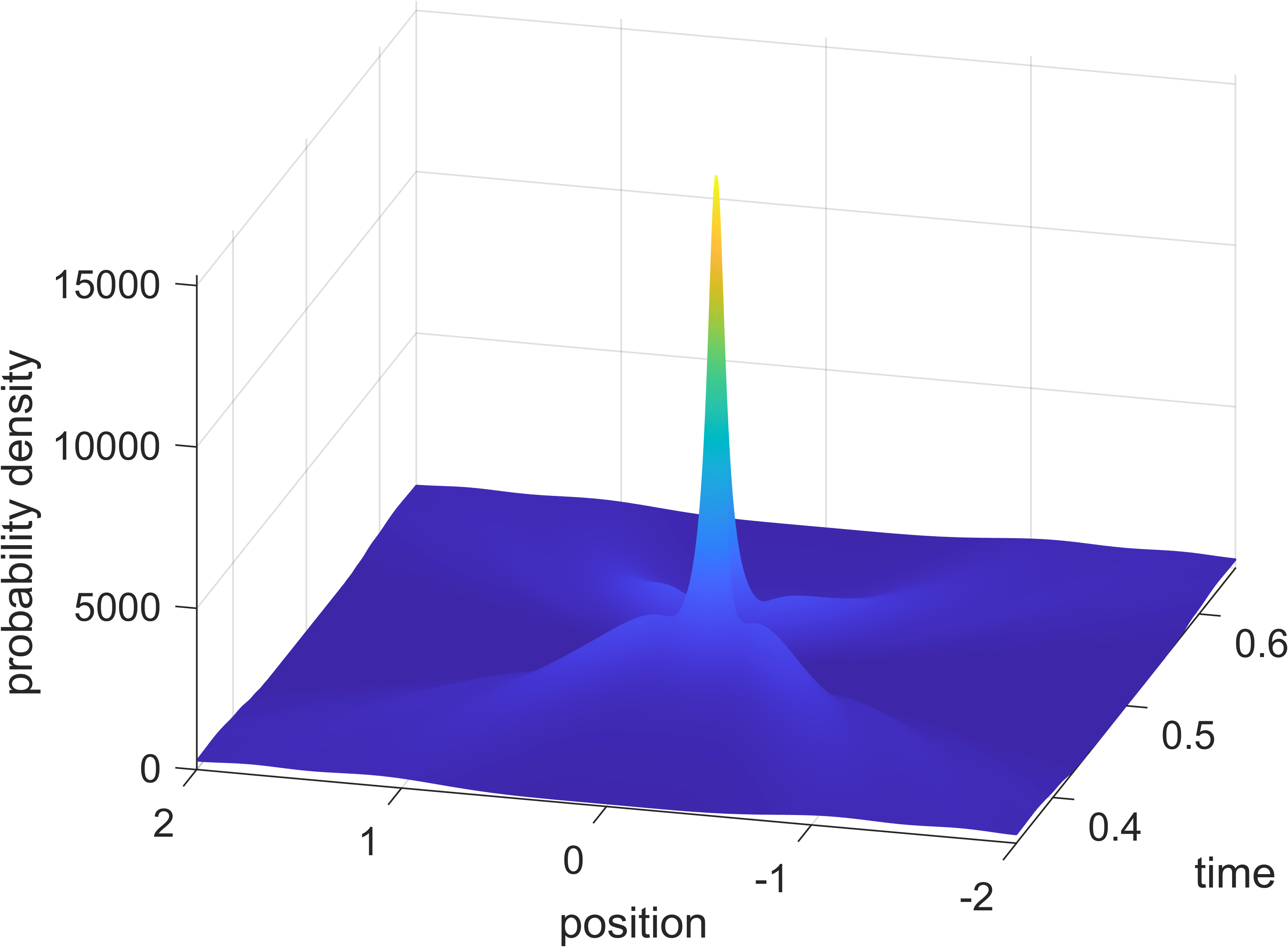}	   
			\caption{}
			\label{fig:nonnodal_b2}
		\end{subfigure}
		\begin{subfigure}[b]{0.47\textwidth}
			\includegraphics[width=\textwidth]{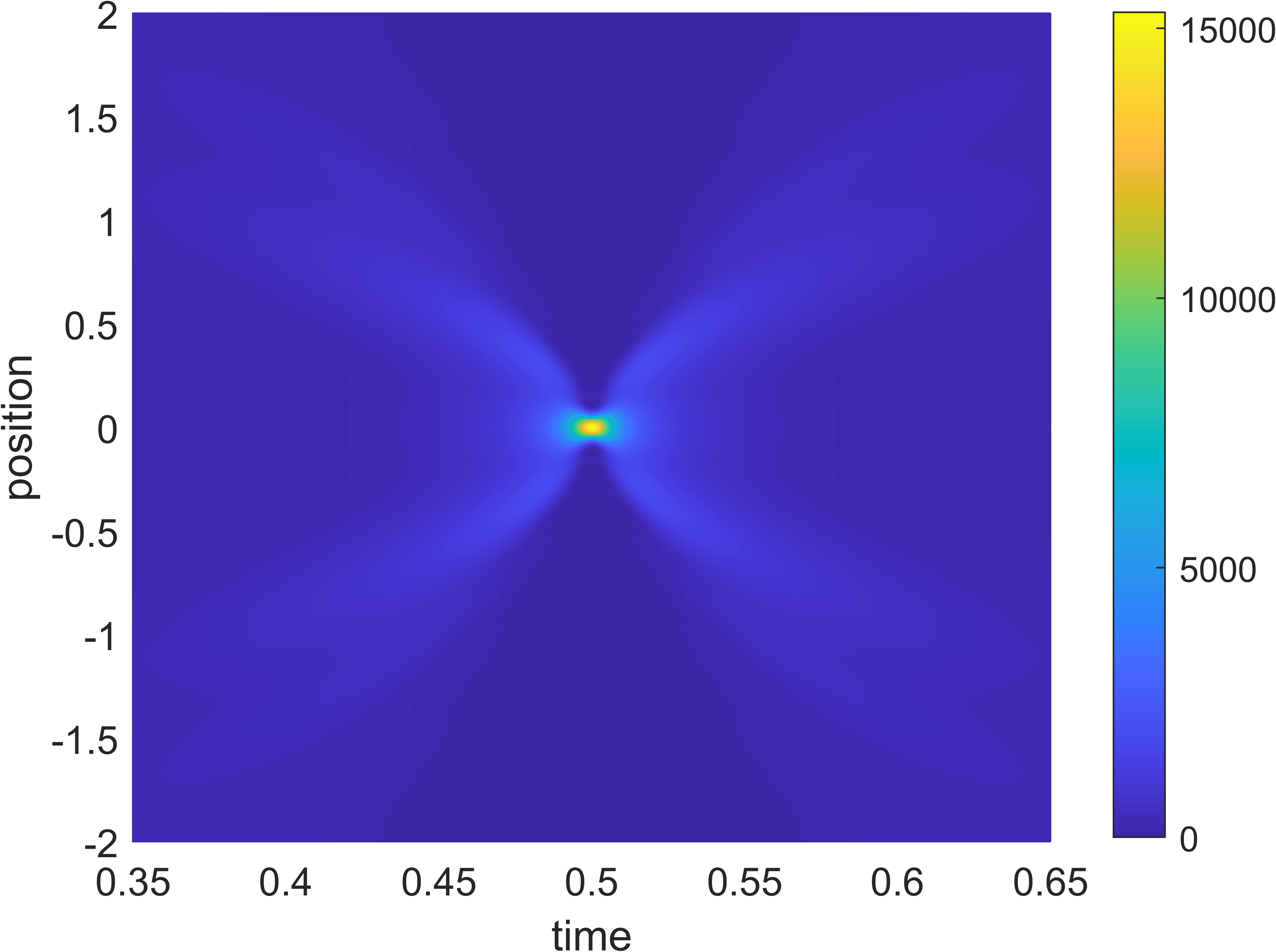}
			\caption{}
			\label{fig:nonnodal_top_b2}
		\end{subfigure}
		\begin{subfigure}[b]{0.47\textwidth}
			\includegraphics[width=\textwidth]{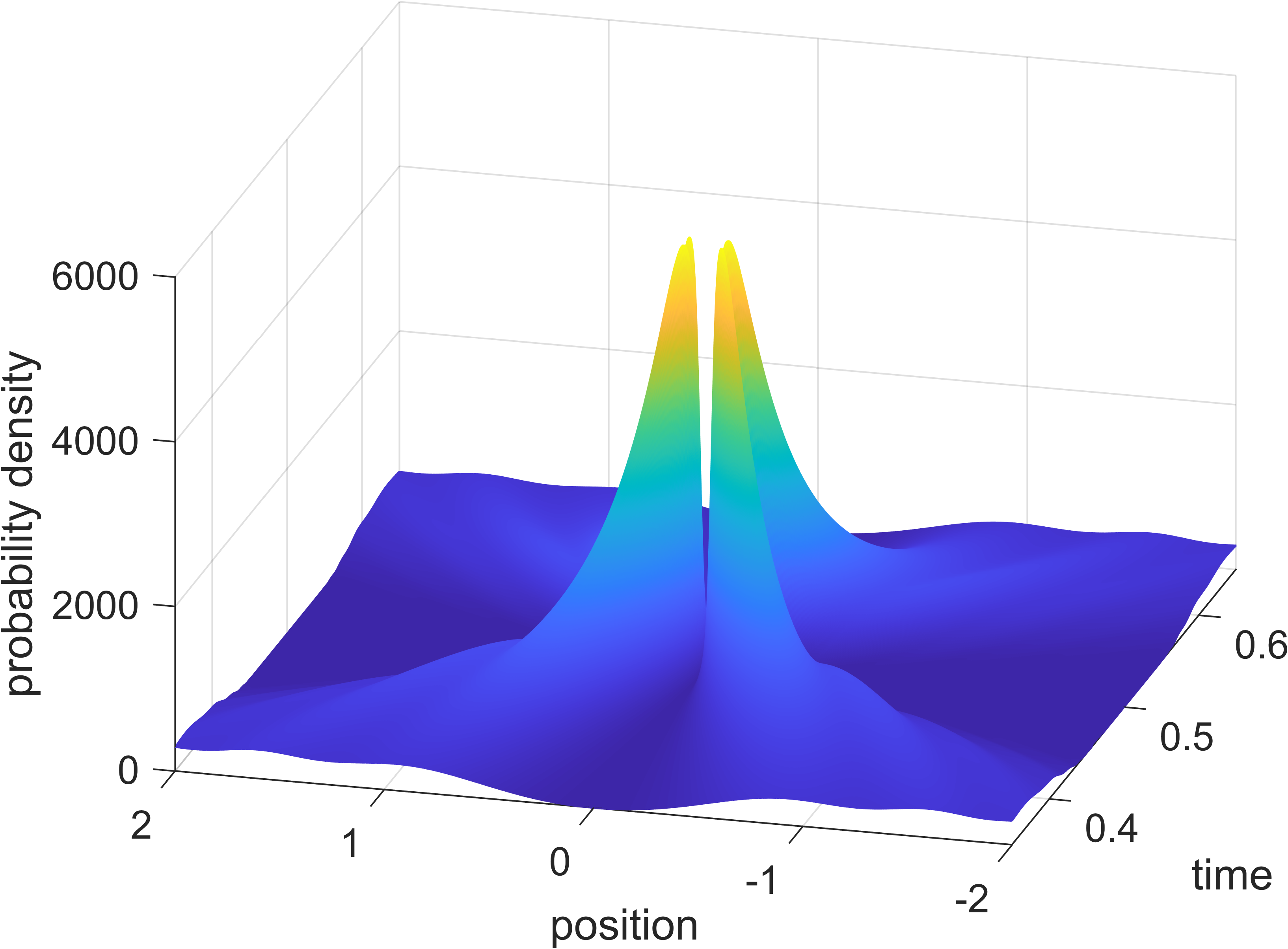}
			\caption{}
			\label{fig:nodal_b2}
		\end{subfigure}
		\begin{subfigure}[b]{0.47\textwidth}
			\includegraphics[width=\textwidth]{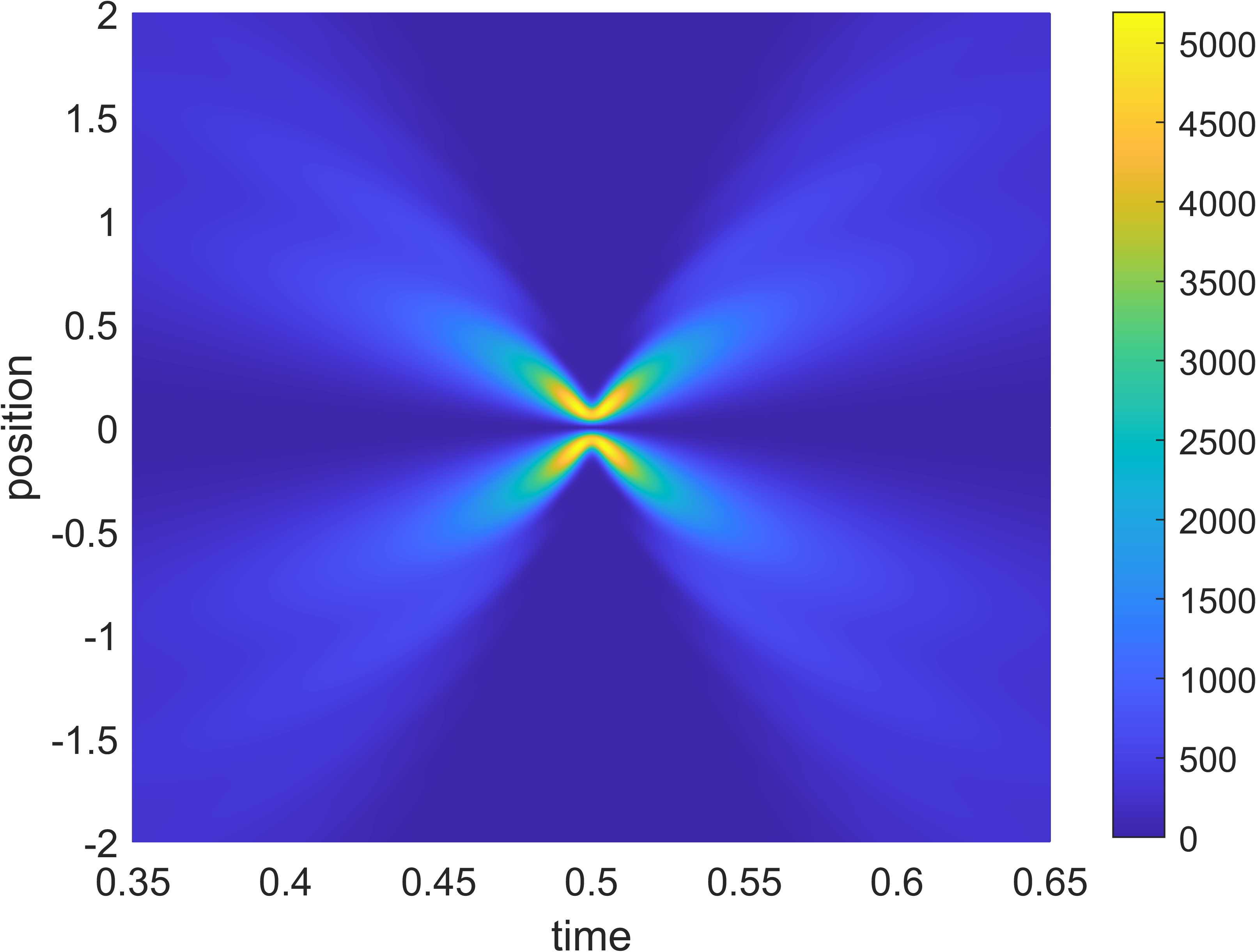}
			\caption{}
			\label{fig:nodal_top_b2}
		\end{subfigure}
		\caption{Time-evolved probability densities for the eigenfunctions  $\varphi_{\tau,0}(q,t)$ $(a,b)$ and  $\varphi_{\tau,1}(q,t)$ $(c,d)$ of the AB operator with $\tau = 0.5 + 0.005i$.  As seen from the top view, the peak of the collapse occurred at $t=\tau_R$. The peaks of (a) and (c) are higher than the peaks of (a) and (c) in Figure 1 indicating that the unitary collapse becomes sharper as $\tau_I$ decreases.}
		\label{fig:TOAevolve-cmplx_t2}
	\end{figure}

\begin{figure}[t!]
		\centering
		\begin{subfigure}[b]{0.47\textwidth}
			\includegraphics[width=\textwidth]{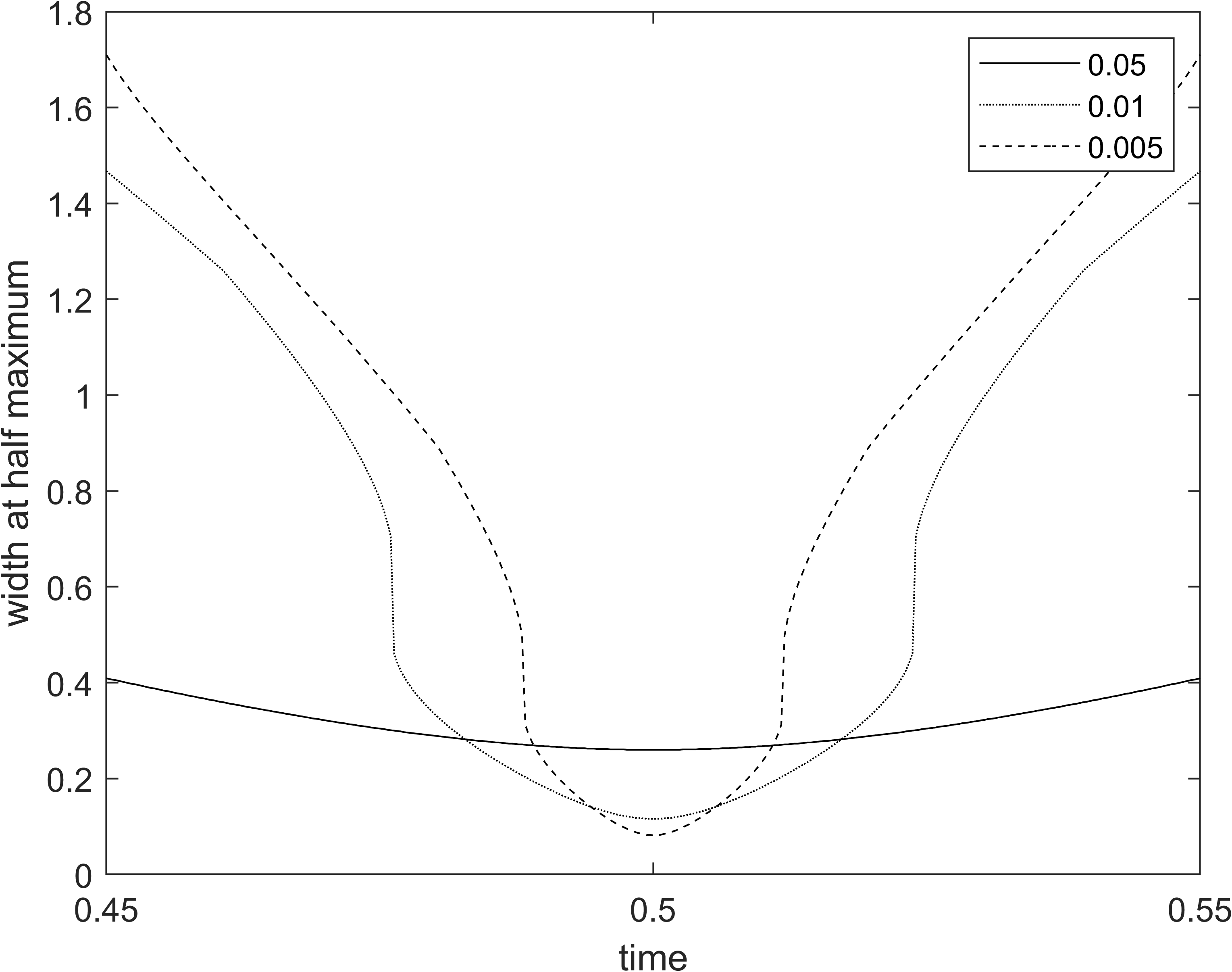}
			\caption{}
			\label{fig:whm_nn}
		\end{subfigure}
		\begin{subfigure}[b]{0.47\textwidth}
			\includegraphics[width=\textwidth]{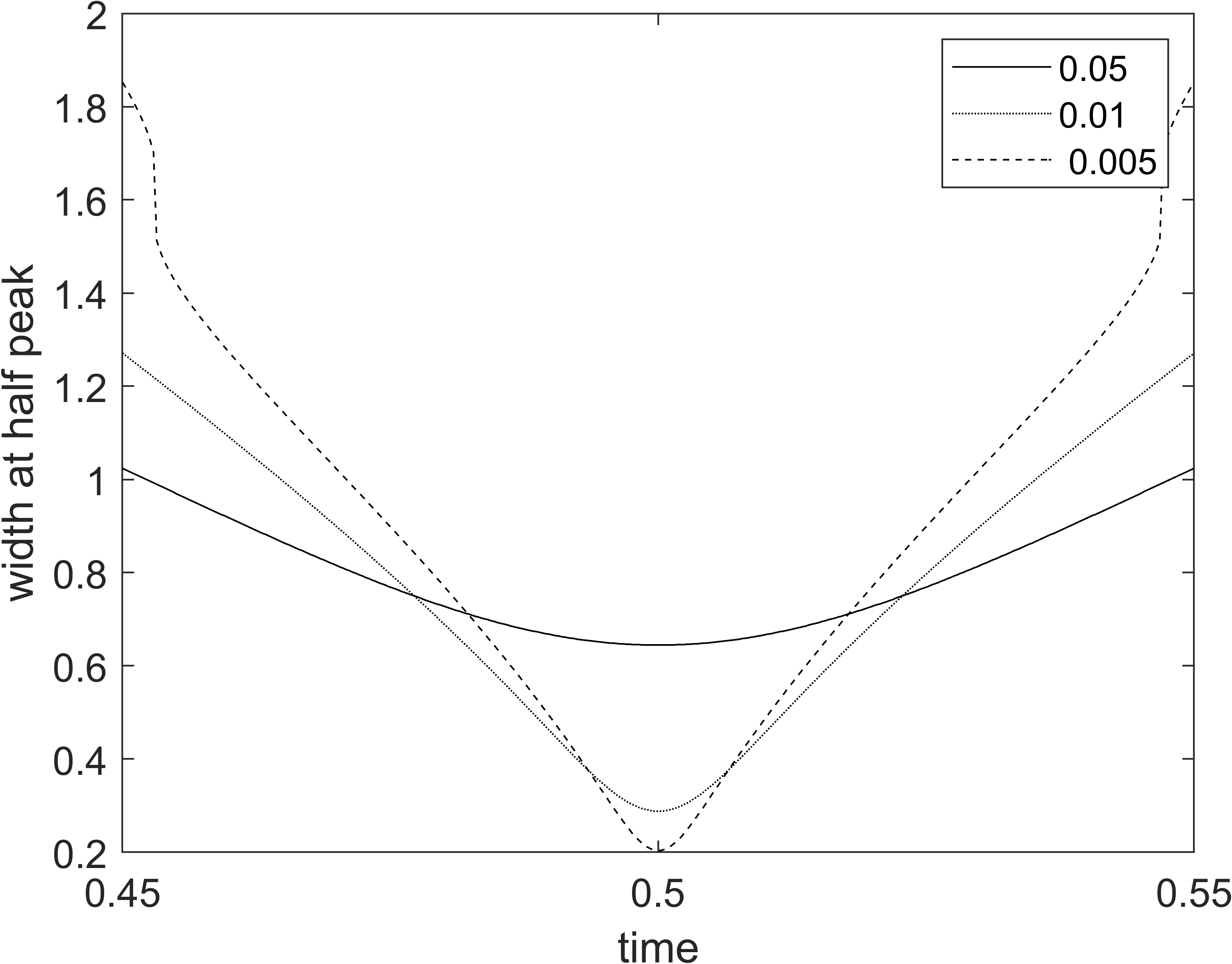}
			\caption{}
			\label{fig:whm_no}
		\end{subfigure}
		\caption{Width-at-half-maximum plot for a.) $|\varphi_{\tau,0}(q,t)|^2$ and b.) $|\varphi_{\tau,1}(q,t)|^2$ for different times $t$ with $\tau = \tau_R + i\tau_I$ where $\tau_R = 0.5$ while $\tau_I$ is varied (see legend). For a given $\tau$, WHM is minimum at $t=\tau_R$. The WHM at $t=\tau_R$ decreases with decreasing $\tau_I$ .}
		\label{fig:TOAeigf_whm}	
	\end{figure}

	We first obtain the time-evolved position probability distribution  $|\varphi_{\tau,n}(q,t)|^2$. For an arbitrary free-particle wavefunction $\psi(p,t=0)$, its time-evolved form in momentum representation is given by $\psi(p,t) = e^{-ip^2 t/(2\mu \hbar)} \psi(p,0)$. It follows from Eqs. \eqref{eq:toa_peigf1} and \eqref{eq:toa_peigf2} that the time evolution of $\varphi_{\tau,n}(p)$ is given by $\varphi_{\tau,n}(p,t) = \varphi_{\tau-t,n}(p,0)$ which is the covariance-under-time-translation property of the TOA eigenfunctions. Note that this property also holds for when $\varphi_{\tau,n}$ is in position representation. Hence, $\varphi_{\tau,n}(q,t)$ is given by
	\begin{equation}\label{eq:TOA_eigf_tevolve}
	\varphi_{\tau,n}(q,t) = \sqrt{\frac{8\tau_I \pi}{|\tau-t|}}\>\Gamma\left(\frac{1}{4}+\frac{n}{2}\right)^{-1}\left[\frac{\mu}{8\hbar |\tau-t|}\right]^{\frac{1}{4}+\frac{n}{2}}q^n \>_1F_1\left(\frac{3}{4}+\frac{n}{2};\frac{1}{2}+n;\frac{-i\mu q^2}{2\hbar (\tau-t)}\right)
	\end{equation}
	Figure \ref{fig:TOAevolve-cmplx_t1} shows the time-evolved probability density $|\varphi_{\tau,n}(q,t)|^2$ for $\tau = 0.5  + 0.01 i$ for the parameter values $\mu = \hbar = 1$. Note that for both eigenfunctions, the position expectation value is always zero since both have symmetry with respect to the origin. The TOA eigenfunctions exhibit the unitary arrival (collapse) property since they evolve according to Schrodinger equation in a way where the spread in the position becomes minimum at a particular point in time. At this instant in time, the probability density is concentrated in the neighborhood of the arrival signifying the arrival of the particle at the arrival point. We see in the top view of Figures \ref{fig:TOAevolve-cmplx_t1} and \ref{fig:TOAevolve-cmplx_t2} that the unitary collapse occurs at $t=\tau_R$, irrespective of the value of $\tau_I$. This is supported in Figure \ref{fig:TOAeigf_whm} of the width-at-half-maximum (WHM) of $|\varphi_{\tau,n}(q,t)|^2$ at times where the  minima is at  $t=\tau_R$ regardless of the value of $\tau_I$.  
	
\begin{figure}[t!]
		\centering
		\includegraphics[width=0.5\textwidth]{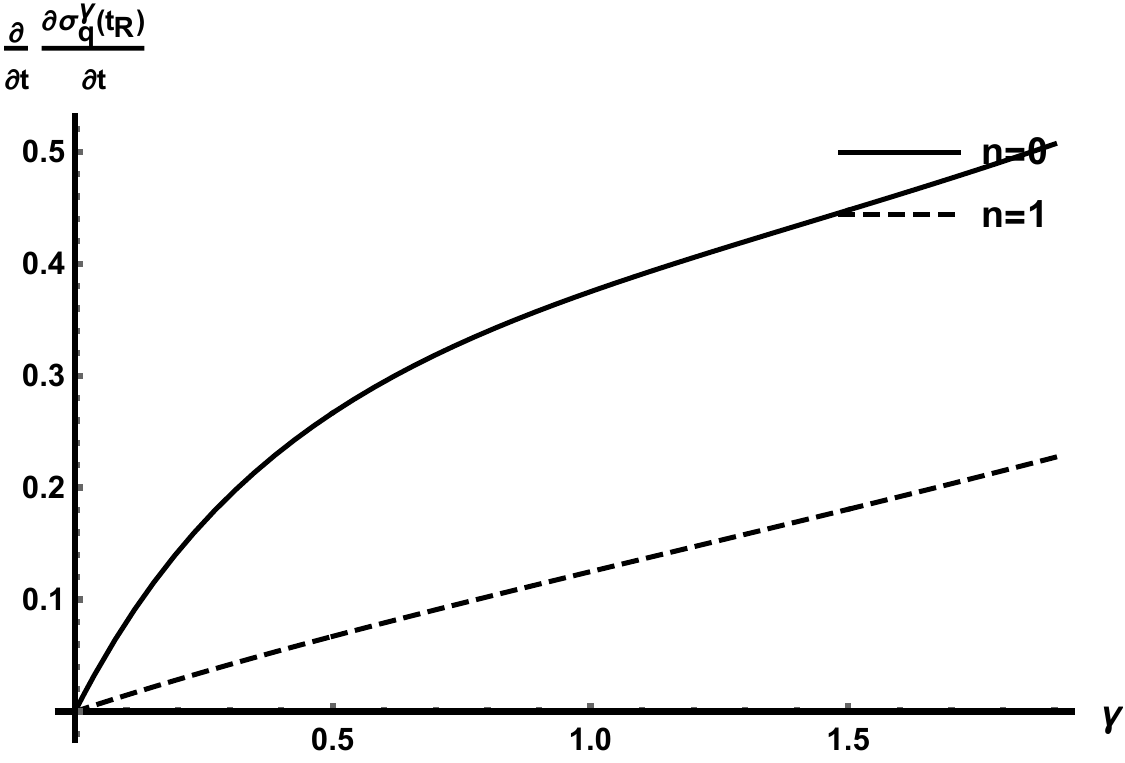}
		\caption{Concavity of $\sigma_q^{\gamma}(t)$ at $t=\tau_R$ for different values of $\gamma$ and for $\tau_I=1$. The concavity is positive for $0<\gamma<2$ which implies that $\sigma_q^{\gamma}(t)$ has a local maximum at $t=\tau_R$. Note that the we can apply an appropriate change of variable to place the dependence of $\tau_I$ outside the integral in Eq. \eqref{eq:spread-mod} so that this result holds true for all $\tau_I>0$.}
		\label{fig:TOA_conc}	
	\end{figure}

	Now, we confirm that the unitary arrival is sharpest at $t=\tau_R$. We accomplish this by showing that the spread in the particle's position is minimum at $t=\tau_R$. The spread is usually described in terms of the position uncertainty. However, $q^2 |\varphi_{\tau,n}(q,t)|^2\sim O(|q|^{-1})$ as $|q|\to\infty$ so that the expectation value  $<\opr{q}^2>$ is infinite, implying that the uncertainty of the position cannot be used to measure the spread of the position. Given that the position expectation value for $\varphi_{\tau,n}(q,t)$ is zero for all $t$, we can describe the spread of $ |\varphi_{\tau,n}(q,t)|^2$ about $q=0$ using a modified variance formula
	\begin{equation}\label{eq:spread-mod}
	\sigma_q^{\gamma}(t) = \int_{-\infty}^{\infty}\mathrm{d}q \> |q|^{\gamma} \left|\varphi_{\tau,n}(q,t)\right|^2
	\end{equation} 
	which exist for any $\gamma \in [0,2)$. The spread $\sigma_q^{\gamma}(t)$ has a local minimum at $t=\tau_R$ if and only if  $\partial_t \sigma_q^{\gamma} = 0$ and $\partial^2_t \sigma_q^{\gamma} > 0$ at $t=\tau_R$. The first requirement $\partial_t \sigma_q^{\gamma}|_{t=\tau_R} = 0$ is straightforward to show since $\partial_t \left[\left|\varphi_{\tau,n}(q,t)\right|^2\right]|_{t=\tau_R} =0$. For the second requirement, we plot the concavity of $\sigma_q^{\gamma}$ at $t=\tau_R$ as a function of $\gamma$. We see in Figure \ref{fig:TOA_conc} that $\partial^2_t \sigma_q^{\gamma}\big|_{t=\tau_R}$ is positive for all $\gamma \in (0,2)$, which implies that the spread $\sigma_q^{\gamma}(t)$ has a local minimum at $t=\tau_R$. Hence, the unitary collapse for $\varphi_{\tau,n}(q,t)$ is sharpest at $t =\tau_R$, irrespective of the value of $\tau_I$.
	
	In \cite{caballar2, galaponPRA2005a}, it was shown that the CTOA operator has two eigenfunction types that have distinct characteristics in their dynamics. The nodal eigenfunction type has the dynamics where the probability density at the arrival point is zero at all times and there are two peaks in the collapse, while nonnodal eigenfunction type has the dynamics where the probability density is nonzero at the arrival point at all times and there is only one peak in the unitary collapse. The eigenfunctions $\varphi_{\tau,n}$ also have this dynamical feature where $\varphi_{\tau,n=0}$ has a collapse with a single peak, while $\varphi_{\tau,n=1}$ has a collapse with two peaks. Therefore, we can associate the odd eigenfunction (n=1) with the nodal type and the even eigenfunction (n=0) with the nonnodal type. The physical importance of these two types of eigenfunctions is discussed in \cite{sombilloAP2016} where they correspond to two types of arrival with the nonnodal type corresponding to arrival with detection, while the nodal type corresponding to arrival without particle detection.
	
	The imaginary part $\tau_I$ determines the sharpness of arrival. We see this by comparing the peaks in Figures \ref{fig:TOAevolve-cmplx_t1} and \ref{fig:TOAevolve-cmplx_t2} where the peak values increase with decreasing $\tau_I$. This is also confirmed in the WHM plot for $\left|\varphi_{\tau,n}(q,t)\right|^2$ in Figure \ref{fig:TOAeigf_whm} where the minimum value of the WHM decreases with decreasing $\tau_I$. This implies that the collapse becomes sharper as $\tau_I$ decreases. We expect that this is sharpest in the limit $\tau_I\to 0^+$ and we infer that in this limit, the space-time distribution is a dirac delta function with support at the space-time point $(x=0,t=\tau_R)$, a result that we will prove in the next section. The imaginary part  $\tau_I$ is also a measure of the energy uncertainty  $\Delta E$ of $\varphi_{\tau,n}(q)$. We see this by computing the energy uncertainty $\Delta E$ of $\varphi_{\tau,n}(q)$ in the momentum representation, 
	\begin{eqnarray} \label{eq:E_unc_tI}
	\Delta E \>  = \tau_I\sqrt{<E^2> - <E>^2} = \sqrt{\frac{\hbar^2}{2\tau_I^2} - \left[\frac{\hbar}{2\tau_I}\right]^2 } = \frac{\hbar}{2\tau_I}
	\end{eqnarray}
	which shows the inverse relationship between $\tau_I$ and $\Delta E$. This suggests that $\Delta E$ determines the sharpness in the arrival of the particle; the unitary collapse is sharper for $\varphi_{\tau,n}(q)$ with higher values of $\Delta E$. In the limiting case $\tau_I \to 0^+$, $\varphi_{\tau,n}(q)$ with real $\tau$ have infinite energy uncertainty and the fact that these are the eigenstates with the sharpest collapse is consistent with the fact that confining the particle in a single location (i.e. arrival at the arrival point) requires a measurement with infinite energy uncertainty. For this reason, real-eigenvalued TOA eigenfunctions can be considered idealized but unphysical TOA eigenstates because these are the eigenstates with definite arrival time at the arrival point but preparing such states requires a process with infinite energy uncertainty. 
	
	It is interesting to note that Eq. \eqref{eq:E_unc_tI} can be written in the form $\tau_I \Delta E = \frac{\hbar}{2}$ but this cannot be interpreted as the minima of the time-energy uncertainty relation because $\tau_I$ is not the uncertainty in the time-of-arrival measurement. What Eq. \eqref{eq:E_unc_tI} means is that there is some kind of time-energy relation where $\tau_I$ is a measure of $\Delta E$ of $\varphi_{\tau,n}(q)$. On the other hand, one might wonder if there is some kind of uncertainty relation between energy and TOA observable for $\varphi_{\tau,n}(q)$ and if this is consistent with the well-known Heisenberg uncertainty relation for energy and time. If the uncertainties are computed from the operators, then $\Delta E$ is given by Eq. \eqref{eq:E_unc_tI} whereas $\Delta T=0$ since $\varphi_{\tau,n}(q)$ is an eigenfunction of $\opr{T}$. Their product is equal to zero and clearly does not follow the time-energy uncertainty principle. This seeming contradiction can be explained by noting that the correct formulation of the uncertainty relation, when uncertainties are computed using operators, is given by $\Delta E \Delta T \geq \frac{1}{2}|<[\opr{H},\opr{T}]>|$. For the relation to hold, $\varphi_{\tau,n}(q)$ must be in the domain of $[\opr{H},\opr{T}]$, which in this case is not satisfied. Hence, there is no energy-time uncertainty relation for TOA eigenstates $\varphi_{\tau,n}(q)$ with complex eigenvalues.
	
	\section{Delta sequence convergence of time-evolved position probability density of the AB operator's eigenfunctions} 
	
	We have seen that the complex-eigenvalued eigenfunctions exhibit unitary collapse, although the collapse is unsharp because it is not a dirac-delta function. Furthermore, the collapse is observed to become sharper as $\tau_I$ decreases, indicating that the unitary collapse becomes a function with singular support for real-eigenvalued eigenfunctions. We now give the proof that this is indeed the case by showing that the function $|\varphi_{\tau,n}(q,t=\tau_R)|^2$ is a dirac delta sequence in the limit $\tau_I \to 0^+$. Our proof involves the use of the following theorem \cite{gel1}.
	
	\begin{theorem}[Gelfand-Shilov]
	    A sequence of functions, $\{f_{\nu}(q)\}$, is a delta-convergent sequence if (1) for any $M>0$ and for $|\alpha|, |\beta| \leq M$, the quantities $|\int_{\alpha}^{\beta} f_{\nu}(q) \mathrm{d}q|$ are bounded by a constant depending only on M, and (2) for any fixed non-vanishing $\alpha$ and $\beta$
	\begin{equation}
	\lim_{\nu\to0^+} \int_{\alpha}^{\beta} \mathrm{d}q \> \> f_{\nu}(q) = \begin{cases}
	0 \qquad ,& \qquad 0<\alpha<\beta \> \text{ or } \> \alpha<\beta<0 \\
	1\qquad ,& \qquad \alpha<0<\beta
	\end{cases}
	\end{equation}
	\end{theorem} 
	
	Consider the sequence of functions $\{f_{\nu = \tau_I}(q) = |\varphi_{\tau,n}(q,\tau_R)|^2\}$. We show that the sequence satisfies the two conditions for a delta sequence in the limit as $\tau_I\to 0$. From Eq. \eqref{eq:TOA_eigf_tevolve}, $|\varphi_{\tau,n}(q,\tau_R)|^2$ is given by
	\begin{equation}
	\left|\varphi_{\tau,n}(q,\tau_R)\right|^2 = \frac{8\pi q^{2n}}{\Gamma\left(\frac{1+2n}{4}\right)^{2}} \left[\frac{\mu}{8\tau_I  \hbar}\right]^{\frac{1}{2}+n} \>_1F_1\left(\frac{3+2n}{4};\frac{1}{2}+n;\frac{-\mu q^2}{2\hbar \tau_I}\right)^2
	\end{equation}

	For the first condition, consider some arbitrary constants $\alpha$, $\beta$ and $M>0$ where $|\alpha|,|\beta|\leq M$. Note that $\left|\int_{\alpha}^{\beta} |\varphi_{\tau,n}(q,\tau_R)|^2 \mathrm{d}q\right| =\int_{\alpha}^{\beta} |\varphi_{\tau,n}(q,\tau_R)|^2 \mathrm{d}q $ so that 
	\begin{equation}
	\left|\int_{\alpha}^{\beta} |\varphi_{\tau,n}(q,\tau_R)|^2 \mathrm{d}q\right| \leq \int_{-\infty}^{\infty} |\varphi^{(n)}_{\tau}(q,\tau_R)|^2 \mathrm{d}q = 1
	\end{equation}
	since $\varphi_{\tau,n}(q,\tau_R)$ for $\tau_I>0$ are all normalized to unity. Therefore, the sequence $\{f_{\nu}(q)\}$ satisfies the first condition for delta-convergent sequence.

	To test for the second condition, we first consider the case where $0<\alpha <\beta$. We have
	\begin{eqnarray}\label{eq:DDS_test1}
	\lim_{\tau_I\to 0^+} \int_{\alpha}^{\beta} \mathrm{d}q \> \> |\varphi_{\tau,n}(q,\tau_R)|^2 &\leq&  \lim_{\tau_I\to 0^+}  \int_{\alpha}^{\infty} \mathrm{d}q \> \> |\varphi_{\tau,n}(q,\tau_R)|^2 \\
	&=& \frac{1}{2} - \lim_{\tau_I\to 0^+} \int_{0}^{\alpha} \mathrm{d}q \> \> |\varphi_{\tau,n}(q,\tau_R)|^2 \nonumber 
	\end{eqnarray}
	The second line follows from the normalization condition and from the symmetry of $|\varphi_{\tau,n}(q,\tau_R)|^2$  about the origin. To evaluate the second term in Eq.\eqref{eq:DDS_test1}, we place the dependence on $\tau_I$ from the integrand into the bounds of the integral through a change of variables $u = \sqrt{\frac{\mu}{2\hbar \tau_I}}q$. The integral becomes
	\begin{eqnarray}
	\int_{0}^{\alpha} \mathrm{d}q \> \> |\varphi_{\tau,n}(q,\tau_R)|^2 = \frac{\pi 4^{1-n} }{\Gamma\left(\frac{1+2n}{4}\right)^{2}} \int_{0}^{\tilde{\alpha}} \mathrm{d}u \> \>  u^{2n} \>_1F_1 \left(\frac{3+2n}{4};\frac{1}{2}+n;-u^2\right)^2
	\end{eqnarray}
	where $\tilde{\alpha} =\sqrt{\frac{\mu}{2\hbar \tau_I}}\alpha$. In the limit as $\tau_I \to 0^+$, $\tilde{\alpha}\to \infty$ so that
	\begin{eqnarray}\label{eq:dds_form1}
	\lim_{\tau_I\to 0^+} \int_{0}^{\alpha} \mathrm{d}q \> \> |\varphi_{\tau,n}(q,\tau_R)|^2 &=& \frac{\pi 4^{1-n} }{\Gamma\left(\frac{1+2n}{4}\right)^{2}} \int_{0}^{\infty} \mathrm{d}u \> \>  u^{2n} \>_1F_1 \left(\frac{3+2n}{4};\frac{1}{2}+n;-u^2\right)^2 \nonumber \\
	&=& \frac{1}{2}
	\end{eqnarray}
	where the last line in Eq. \eqref{eq:dds_form1} follows from the integral formula (see Appendix for the proof)
	\begin{equation}\label{eq:mtmc_intform1}
	\int_0^{\infty} \> \mathrm{d}q \> q^{2n} \>\left|_1F_1\left(\frac{n}{2}+\frac{3}{4};\frac{1}{2}+n;-cq^2\right)\right|^2 = \frac{4^{-1+n}|c|^{1/2-n}}{2\pi \text{Re}(c)}\Gamma\left(\frac{1}{4}+\frac{n}{2}\right)^2
	\end{equation}
	which holds for $Re(c)>0$. Hence,  $\lim_{\tau_I\to 0^+} \int_{\alpha}^{\beta} \mathrm{d}q \> \> |\varphi_{\tau,n}(q,\tau_R)|^2=0$ for $0<\alpha<\beta$. 
	
	For $\alpha <\beta<0$, $ |\varphi_{\tau,n}(q,\tau_R)|^2 $ is an even function, therefore we can apply a change of variable $q\to-q$. We obtain
	\begin{eqnarray}
	\lim_{\tau_I \to0^+} \int_{\alpha}^{\beta} \mathrm{d}q \> \>  |\varphi_{\tau,n}(q,\tau_R)|^2 =  \lim_{\tau_I\to0^+} \int_{|\beta|}^{|\alpha|} \mathrm{d}q \> \>  |\varphi_{\tau,n}(q,\tau_R)|^2
	\end{eqnarray}
	which is equal to zero for any $|\alpha|$ and $|\beta|$. Finally, for $\alpha<0<\beta$, we obtain the following result
	\begin{eqnarray}
		\lim_{\tau_I \to0^+} \int_{\alpha}^{\beta} \mathrm{d}q \> \> |\varphi_{\tau,n}(q,\tau_R)|^2 &=& 	\lim_{\tau_I \to0^+}\left[ \int_{-\infty}^{\infty} \mathrm{d}q \> \> |\varphi_{\tau,n}(q,\tau_R)|^2 \right. \nonumber \\ && \decspace \decspace \decspace \left.- \int_{-\infty}^{\alpha}  \mathrm{d}q \> \> |\varphi_{\tau,n}(q,\tau_R)|^2 - \int_{\beta}^{\infty} \mathrm{d}q \> \> |\varphi_{\tau,n}(q,\tau_R)|^2 \right] \\
	&=& 1 \nonumber
	\end{eqnarray} 
	Hence, we have proven that $|\varphi_{\tau,n}(q,\tau_R)|^2$ forms a dirac delta sequence as $\tau_I \to 0^+$,
	\begin{equation}
	\lim_{\tau_I\to0^+} |\varphi_{\tau,n}(q,\tau_R)|^2 = \delta(q)
	\end{equation}

	\section{Discussion and Conclusions}
	
	In this paper, we have given a mathematical proof that the nonnormalizable, real-eigenvalued eigenfunctions of the Aharonov-Bohm operator defined in the entire real line represents a quantum state with definite arrival time at the arrival point. The crucial part of the proof is the use of the complex-eigenvalued eigenfunctions of the AB operator, states that reside outside of the domain of the hermitian $\opr{T}$ but are physically realizable (e.g. reside in the hilbert space) and are eigenfunctions of the non-hermitian $\opr{T}$. These eigenfunctions can be considered as legitimate TOA eigenfunctions, since they possess the unitary arrival property, only that arrival is not sharp because it is not a Dirac delta function at the eigenvalue. The real part of the eigenvalue is the time at which the peak of the unitary collapse occurs, whereas the imaginary part of the eigenvalue is a measure of the energy uncertainty of the state and determines the sharpness of the unitary collapse. We have shown that the time-evolved TOA position probability distribution evaluated at the time equal to the real part of the eigenvalue is a dirac delta sequence in the limit as the imaginary part of the eigenvalue approaches zero, proving that the real-eigenvalued eigenfunctions of the AB operator are states with definite arrival time at the arrival point. This result gives us more confidence in representing the free-particle time-of-arrival observable with the AB operator, since the operator's properties are consistent with what is expected of the time-of-arrival observable. This also shows that the hermitian $\opr{T}$ is the idealized representation of the free TOA observable having eigenfunctions that have a definite arrival time while its non-hermitian $\opr{T}$ can be considered as its unsharp version where its eigenfunctions have unsharp arrival.
	
	We have seen that nodal and nonnodal types give distinct types of delta sequence with the same limit delta function. The delta sequence obtained from the nodal type is an example of the type of delta sequence first demonstrated in \cite{galaponJPA2009} where all its elements vanish at the arrival point. The question now arises whether there is any difference in a dirac delta function obtained from the nodal type from that obtained from the nonnodal type.This is interesting because mathematically a dirac delta function is just one object defined by its point support, and in quantum mechanics, this represents a particle with an exact location. However, our result seems to indicate the need for distinction because the two eigenfunction types are different from each other. 
	
	In \cite{sombilloAP2016}, the significance of the two TOA eigenfunction types is investigated through the comparison of the TOA distribution profile with the spatio-temporal profile of the particle's initial wavefunction. The nonnodal type is associated with arrival with detection because its contribution to the TOA distribution profile matches $|\psi(x,t)|^2$, while the nodal type is associated with arrival without detection because its contribution to the TOA profile has a maximum at the points where $|\psi(x,t)|^2=0$. The concept of arrival with and without detection might distinguish the dirac delta function based on the properties of its delta sequence representation. A particle located exactly at the arrival point, that is, $|\psi(q)|^2 = \delta(q)$, does not automatically mean that the particle will be detected at the point. A dirac delta function coming from the nodal type could imply that the particle is actually located in the vicinity of the point. We note that the possible multiple interpretations of a dirac delta function should not be a cause of concern as to how it is used in quantum mechanics because these states are not physically realizable as can only be obtained in measurements with infinite energy spread.

	\section*{Appendix}
	\subsection*{Proof of Equation \eqref{eq:mtmc_intform1}}
	We prove the integral formula Eq. \eqref{eq:mtmc_intform1}. For $0<Re(a)<Re(b)$, the hypergeometric function $ _1F_1(a;b;z)$ has the integral representation along the real line given by \cite[p. 326]{NIST_Olver}
	\begin{equation}\label{eq:1F2_intrep}
	\> _1F_1(a;b;z) = \frac{\Gamma(b)}{\Gamma(b-a)\Gamma(a)}\int_0^1 \frac{t^{a-1} e^{zt}}{(1-t)^{1+a-b}} \mathrm{d} t
	\end{equation}
	Applying this gives us
	\begin{eqnarray}
	&& \int_0^{\infty} \> \mathrm{d}q \> q^{2n} \>\left|_1F_1\left(\frac{n}{2}+\frac{3}{4};\frac{1}{2}+n;-cq^2\right)\right|^2 =\left(\frac{4\Gamma(n + \frac{1}{2})}{(2n - 1)\Gamma(\frac{n}{2} - \frac{1}{4})^2}\right)^2 \nonumber \\ 
	&&\times \int_0^{\infty} \mathrm{d}q\> q^{2n}  \int_0^1 \int_0^1 \mathrm{d}t\mathrm{d}t' \frac{(tt')^{\frac{2n-1}{4}} e^{-q^2(ct+c^*t')}}{((1-t)(1-t'))^{{\frac{5-2n}{4}}}} 
	\end{eqnarray}
	We perform the integration with respect to $q$ first  using the integral formula  $\int_0^{\infty} \mathrm{d}q q^{2n}e^{-\alpha q^2} = \frac{1}{2}\alpha^{-1/2-n} \Gamma\left(\frac{1}{2}+n\right)$ for $Re(\alpha)>0$. We then evaluate the double integral by noting that it can be written as a beta function 
	\begin{equation}
	B(z_1,z_2) = (1-a)^{z_2} \int_0^1 \frac{(1-u)^{z_1-1}u^{z_2-1}}{(1-au)^{z_2+z_1}}\mathrm{d}u
	\end{equation}
	which holds for any $a\in \mathcal{C}, a\neq 1$. Evaluating the double integral,
	\begin{eqnarray}
	\int_0^1 \int_0^1 \mathrm{d}t\mathrm{d}t' \frac{(tt')^{\frac{2n-1}{4}}  (ct+c^*t')^{-\frac{1+2n}{2}} }{((1-t)(1-t'))^{\frac{5-2n}{4}}} =  \int_0^1 \mathrm{d}t  \frac{ B\left(\frac{n}{2}+ \frac{3}{4}, \frac{n}{2}-\frac{1}{4}\right)  c  |c|^{-\frac{2n+3}{2}}}{(1-t)^{\frac{5-2n}{4}}(1+\frac{ct}{c^*})^{\frac{2n+3}{4}}} \nonumber \\
	= B\left(\frac{2n+3}{4}, \frac{2n-1}{4}\right)  B\left(\frac{2n-1}{4},1\right) \frac{c  |c|^{-\frac{2n+3}{2}}}{1+\frac{c}{c^*}}
	\end{eqnarray}
	Writing the beta function in terms of the gamma function, $B(z_1,z_2) = \frac{\Gamma(z_1)\Gamma(z_2)}{\Gamma(z_1+z_2)}$, we obtain Eq. \eqref{eq:mtmc_intform1}. The integral representation given by Eq. \eqref{eq:1F2_intrep} does not hold for $n=0$ case. However, we can apply analytic continuation to Eq. \eqref{eq:mtmc_intform1} to show that it holds for $n>-1/2$.


\begin{thebibliography}{99}
		
		\bibitem{muga2} J. G. Muga  \& C. R. Leavens  {\it  Phys. Rep.} {\bf 338} 353, (2000).  
		
		\bibitem{leavens} C.R. Leavens {\it Phys. Rev. A} {\bf 58} (1998) 840. 
		
		\bibitem{Busch_pla} P. Busch, M. Grabowski, P.J. Lahti, Phys. Lett. A 191 (1994) 357.
		
		\bibitem{toa4} N. Grot, C. Rovelli, and R. Tate, {\it Phys.Rev. A} {\bf 54} 467 (1996). 
		
		\bibitem{giannitrapani1997} R. Giannitrapani,  {\it Int. J. Theor. Phys.}, {\bf 36}, 1575-1584 (1997) 
		
		\bibitem{delgado1997PRA} V. Delgado, J.G. Muga, {\it Phys. Rev. A} 56 (1997) 3425.
		
		\bibitem{delgado1998PRA} Delgado, V. {\it Phys. Rev. A} 57.2 (1998): 762. 
		
		
		\bibitem{leon1997JPA} J. Leon, {\it J. Phys. A} 30 (1997) 4791.
		
		\bibitem{Muga1998} J.G. Muga, C. R. Leavens and  J.P. Palao, {\it Phys. Rev. A}, {\bf 58} 4336 (1998) 
		
		\bibitem{mug21} A. D. Baute, I. Igusquiza \& J. G. Muga {\it Phys. Rev. A} {\bf 64} 012501, (2001). 
		
		
		\bibitem{kijowski1999PRA} J. Kijowski, {\it Phys. Rev. A} 59 (1999) 897. 
		
		\bibitem{mug5} I. L. Egusquiza \& J. G. Muga {\it Phys. Rev. A} {\bf 62} 032103 (2000). 
		
		\bibitem{caballar2} E. Galapon, R. Caballar  and R. Bahague, {\it Phys. Rev. Lett.}, {\bf 93} 180406 (2004). 
		
		\bibitem{galaponJMP2004} E.A. Galapon, {\it J. Math. Phys.} {\bf 45}, 3180 (2004). 
		
		\bibitem{galaponPRA2005a}  E. Galapon, R. Caballar and R. Bahague, {\it Phys. Rev. A}, {\bf 72} 062107 (2005). 
		
		
		\bibitem{galaponPRA2005b} E.A. Galapon, F. Delgado, J.G. Muga, and I. Egusquiza, {\it Phys. Rev. A} {\bf 72}, 042107 (2005).
		
		
		\bibitem{villanuevaJPA2008} E.A. Galapon and A. Villanueva, {\it J. Phys. A: Math. Theor.} {\bf 41}  455302 (2008). 
		%
		
		
		\bibitem{villanuevaPRA2010} A.D. Villanueva \& E.A. Galapon {\it Phys. Rev. A} {\bf 82}, 052117 (2010). 
		
		\bibitem{galaponPRSA2009} E.A. Galapon, {\it Proc. Roy. Soc. Lond. A}  {\bf 465}, 71-86 (2009). 
		
		\bibitem{galaponPRA2009} E.A. Galapon, {\it Phys. Rev. A} {\bf 80}, 030102R (2009). 
		
		%
		\bibitem{sombilloJMP2012} D.L.B. Sombillo and E.A. Galapon, {\it J. Math. Phys.} {\bf 53}, 043702 (2012). 
		
		\bibitem{sombilloAP2016} D.L.M. Sombillo  and E.A. Galapon, {\it Ann. Phys.} {\bf 364},  261–273 (2016). 
		
		\bibitem{EAGJJM2018} E.A. Galapon  and J.J.P. Magadan, {\it Ann. Phys.} {\bf 397},  278–302 (2018). 
		
		\bibitem{DAPJJMCAEAG2024} D.A.L. Pablico, J.J.P. Magadan, C.A.L. Arguelles,  and E.A. Galapon, {\it Phys. Lett. A} {\bf 523},  129778 (2024). 
		
		
		\bibitem{das_SciRep_2019} S. Das, D. Dürr. {\it Sci. Rep.} 9, 2242 (2019) 
		
		\bibitem{stopp_NJPhys_2021} Stopp, F., Ortiz-Gutiérrez, L., Lehec, H., Schmidt-Kaler, F. (2021). {\it N. J. Phys.} 23(6), 063002. 
		
		\bibitem{drezet_Symmetry_2024} A. Drezet (2024). {\it Symmetry}, 16(10), 1325. 
		
		\bibitem{ayatollah_CommunPhys_2023} Ayatollah Rafsanjani, A., Kazemi, M., Bahrampour, A. et al. {\it Commun. Phys.} {\bf 6}, 195 (2023). 
		
		\bibitem{ayatollah_SciRep_2024} Ayatollah Rafsanjani, A., Kazemi, M., Hosseinzadeh, V. et al. {\it Sci. Rep.} {\bf 14}, 3615 (2024) 
		
		\bibitem{das_AnnPhys_2025} Das, S., Deckert, D. A., Kellers, L., Krekels, S., Struyve, W. (2025). {\it Ann. Phys.}, 170054. 
		
		\bibitem{galaponPRL2012} E.A. Galapon, {\it Phys. Rev. Lett.} {\bf 108}, 170402 (2012). 

        \bibitem{DAPEAG2020} D. A. L. Pablico and E.A. Galapon (2020). {\it Phys. Rev. A}, 101(2), 022103.
		
		\bibitem{PCFDAPEAG2024_2} P.C. Flores, D.A. Pablico, E. Galapon,  {\it Phys. Rev. A} {\bf 110}, 062223 (2024)
		
		\bibitem{PCFDAPEAG2024_1} P.C. Flores, D.A. Pablico, E. Galapon,  {\it EPL} {\bf 145}, 65002 (2024)
		
		\bibitem{Sainadh2019} U.S. Sainadh, H. Xu, X. Wang et al., {\it Nature} {\bf 568}, 75-77 (2019)
		
		\bibitem{Aharonov1961} Y. Aharonov and D. Bohm, {\it Phys. Rev.}, {\bf 122} 1649  (1961). 
		
		\bibitem{paul1962} H. Paul, {\it Annalen der Physik}, 464(5‐6), 252-261. (1962). 
		
		\bibitem{kijowski} J. Kijowski, {\it Rep. Math. Phys.} {\bf 6}, 362 (1974).  
		
		\bibitem{galaponJPA2009} E.A. Galapon, {\it J. Phys. A: Math. Theor.} {\bf 42} 175201 (2009). 
		
		\bibitem{wolfram1F1} https://functions.wolfram.com/HypergeometricFunctions/Hypergeometric1F1/06/02/0003/
		
		\bibitem{NIST_Olver} F.W. Olver, D.W. Lozier, R.F. Boisvert, and C.W. Clark, {\it NIST Handbook of Mathematical Functions}, 1st ed. (Cambridge University Press, New York, NY, USA, 2010)
		\bibitem{gel1} I. M. Gel'fand \& G. E. Shilov  {\it Generalized Functions: Properties and Operations} vol. 1. Academic Press (1964).
	\end{thebibliography}
\end{document}